\documentclass[a4paper,11pt]{article}
\pdfoutput=1

\usepackage{jinstpub} 

\usepackage{lineno}
\usepackage{ulem}

\title{\boldmath Diagnostics of hot electrons leaving the ECR plasma sustained by the high-power gyrotron}

\author[a,b,1]{E. M. Kiseleva,\note{Corresponding author.}}
\author[a, b]{I. V. Izotov,}
\author[a,b]{V. A. Skalyga}

\affiliation[a]{Institute of Applied Physics, Russian Academy of Sciences (IAP RAS), \\ Nizhny Novgorod, Ulyanova St., 46, Russia}
\affiliation[b]{Lobachevsky State University of Nizhny Novgorod - National Research University (UNN), \\ Nizhny Novgorod, Gagarina Av., 23, Russia}

\emailAdd{kiseleva@ipfran.ru}

\abstract{Energy distribution of electrons in the plasma sustained by the electron-cyclotron resonance (ECR) discharge has a complicated shape as a function of various parameters that still remains unknown. Meanwhile, it is an important plasma characteristic. Some methods and approaches give the possibility to estimate or measure the properties of the distributions of the electrons lost from the plasma. One of them, similar to the ion mass spectrometry \cite{lit8}, was used in this work to obtain such distributions in the non-classical continuous ECR ion source with high (up to $ 50-100$ W/cm$ ^ 3 $) energy input for the first time along with bremsstrahlung spectra. For certain parameters, a threshold-like regime was discovered, which comprised of the bursts of energetic electrons, bremsstrahlung and, supposedly, the development of kinetic instabilities.}


\begin{document}
\maketitle
\flushbottom

\section{Introduction}

Sources of plasma operating at low pressures are frequently found in elementary particles physics as ion sources in accelerators. A great number of such facilities is based on an electron cyclotron resonance (ECR) discharge, which allows effective heating of electrons in magnetic field, if the electron gyrofrequency equals the microwave radiation frequency. 

ECR ion sources have been in great demand over the past decades as they provided effective generation of multicharged ions. However, the requirements for such facilities are constantly getting stronger to fulfill the needs and boost the performance of modern accelerators. One of the main development trends at the present time is the increase in the total ion current and the beam charge maintaining high beam quality. Understanding the physics of the processes occurring in the plasma of an ECR discharge is crucial not only to satisfy the scientific interest, but also with the aim of performing specific improvements and applications.

In most classical ECR ion sources, the plasma is confined in the so-called collisionless mode \cite{Pastukhov}, which can be characterized by high average energy of electrons (mostly transverse) and their long lifetime in the trap. This peculiarity gives the opportunity for increasing the ionization rates of multicharged ions (MCI) in the optimal regime and thus to obtain a high average charge of ions in the plasma and, consequently, the beam. However, the long lifetime of electrons means that the losses of particles, including ions, are small, which leads to low extracted beam current. At the same time, there is another type of ECR ion sources, in which plasma confinement is realized in the so-called quasi-gasdynamic \cite{gasd}, or collisional, mode. Such facilities are characterized by the high plasma density, the low electron lifetime and, therefore, the large total value of the current, with a drawback of the low average ion charge in the beam.

The peculiarities of the ECR heating make the energy distribution of particles in the discharge extremely complicated. However, the knowledge of this function is essential for various reasons. First of all, it is needed for precise calculations. The main mechanism of MCI formation in the ECR plasma of an ion source is stepwise ionization by direct electron impact that occurs when an energetic electron, colliding with an atom, kicks out other electrons from different energy levels. For the typical plasma parameters of $N_e \approx 10^{11}$-$10{12} cm^{- 3}$, <$E$>$\approx  1$ - $10$ keV, other ionization mechanisms are usually insignificant. To calculate the ionization rates, and, consequently, the dynamics of the density of all ions, it is necessary to know the electron energy distribution function (EEDF): $k_i = \int v \sigma(\varepsilon) f(\varepsilon) d\varepsilon$, where $v$ and $\varepsilon$ are the electron velocity and energy, correspondingly, $\sigma(\varepsilon)$ is the ionisation cross-section, $f(\varepsilon)$ is the electron energy distribution function. With these data, it is possible to evaluate the efficiency of the confinement and ionization processes. As a rule, well-known and simple forms of distribution functions are used for calculations: Maxwellian, sometimes bi- and tri-Maxwellian. However, the EEDF of the plasma of an ECR discharge can differ significantly from the Maxwellian distribution and still remains unknown.

Furthermore, the knowledge of the EEDF is necessary for the general understanding of the physical processes occurring in the ECR plasma. One of the specific features of the ECR discharge is that, under certain conditions, a nonequilibrium fraction of electrons appears with the energy noticeably higher than average. This group of particles may make up only a small part of all electrons, but still strongly influence the processes occurring in the plasma, due to the significant amount of energy stored in it. For example, the presence of an unstable high-energy fraction of electrons can  worsen the efficiency of the ECR confinement as a result of the development of kinetic instabilities  \cite{lit2b}, thus dramatically reducing the performance of the ion source. This leads to periodic destruction of the ambipolar potential that traps ions, which is the cause of a significant decrease in the average ion charge, oscillations of the ion beam current \cite{osc}, the appearance of a large amount of bremsstrahlung radiation, and, last but not least, large thermal loads on superconducting magnets. 

Understanding of the EEDF formation is also essential in the theory of interaction of microwave radiation with plasma. Such research is undertaken with the intention of obtaining the possibility to control the energy distribution and modify ion composition in the extracted beam for the optimal performance without significant changes in the facility. For example, there were experiments \cite{tuning}, in which, due to the fine tuning of the parameters of the system (apparently, it was the EEDF), the MSI current increased several times during the transition to a quasi-unstable plasma confinement regime with no modifications made to the system.

One of the most effective ways to increase the ECR ion sources performance in recent years has been the use of two-frequency plasma heating. It was shown in experiment \cite{lit2b} that for the same total input power, one can obtain significant improvements in the characteristics of the extracted beam by increasing the plasma stability at a higher input energy or the strength of the magnetic field in the trap. The theory described above has shed light to the explanation of this effect. However, such a model still doesn't fully explain the distributions that are seen in the experiment.

Some theoretical assumptions exist that might explain the effects obtained experimentally. To comment on this, it is necessary to mention the quasilinear theory of electron-cyclotron interaction in a mirror magnetic field. A comprehensive model was described in \cite{control}. It explains the reason for the appearance of a hot fraction of electrons in the plasma of an ECR discharge and the efficiency of two-frequency heating. The main theses of the idea can be explained as follows: with the plasma ECR heating by monochromatic radiation, a quasi-one-dimensional in the phase space EEDF is formed. The distribution, which is obtained in such a process, can be unstable, which leads to the excitation and amplification of some parasitic EM waves. In the two-frequency regime, however, this diffusion takes place in all directions in the phase space, thus "smearing" the EEDF, which leads to a significant decrease in the hot electron fraction and the threshold for the development of kinetic instabilities.

Apart from obtaining the real EEDF, there are some ways to estimate the characteristic energies of electrons inside the plasma chamber. Most commonly, the bremsstrahlung spectra are used. This type of radiation is arising from the scattering of electrons of the discharge in an electric field of the atoms and ions, in particular, on the walls of the plasma chamber. However, such method provides only qualitative information on the efficiency of the electron heating due to the ambiguity of the relationship between the EEDF and the bremsstrahlung spectrum.

This work is aimed to present the energy distributions of the energetic electrons lost from the ECR plasma in the CW discharge with high energy input compared to the other existing facilities and to discuss the perspectives of the presented method of diagnostics combined with the bremsstrahlung spectra for the development of the new and only way to restore the real EEDF inside the plasma. First experiments of such kind in the CW mode were undertaken in \cite{lit2}.

\section{Experimental setup}
\begin{figure*}\centering
	\vspace{0cm} 
    \includegraphics[width=120mm]{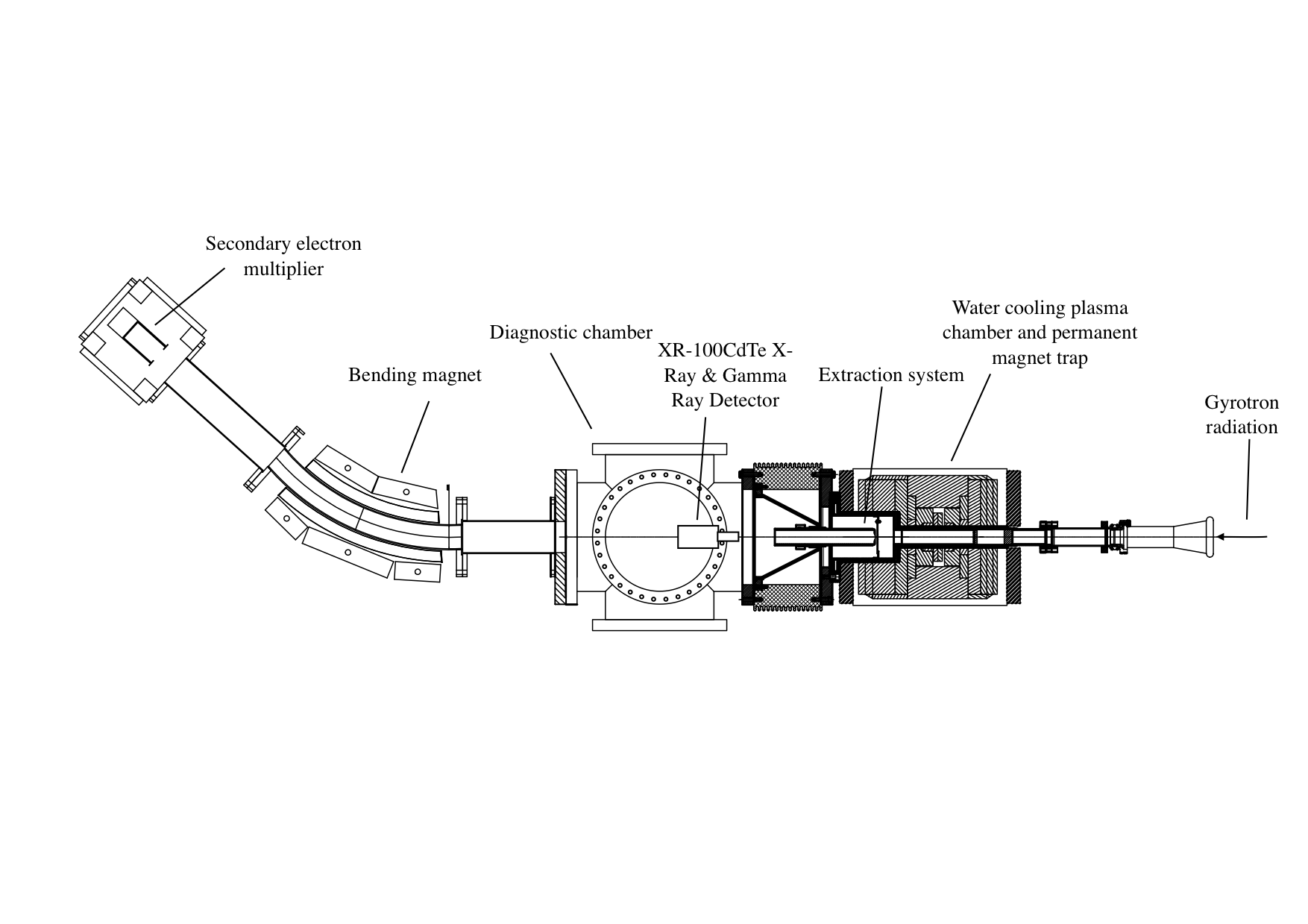}
	\caption{\label{fig2} The experimental scheme}
	\vspace{0cm} 
\end{figure*}

The results presented in this work were obtained on the Gasdynamic Ion Source for Multipurpose Operation facility, a gasdynamic ECR source with a high specific energy input constructed at the IAP RAS \cite{lit1}. The scheme of the facility is presented in Fig.\ref{fig2}.

\begin{figure*}[t]
\centering
\includegraphics[width=80mm]{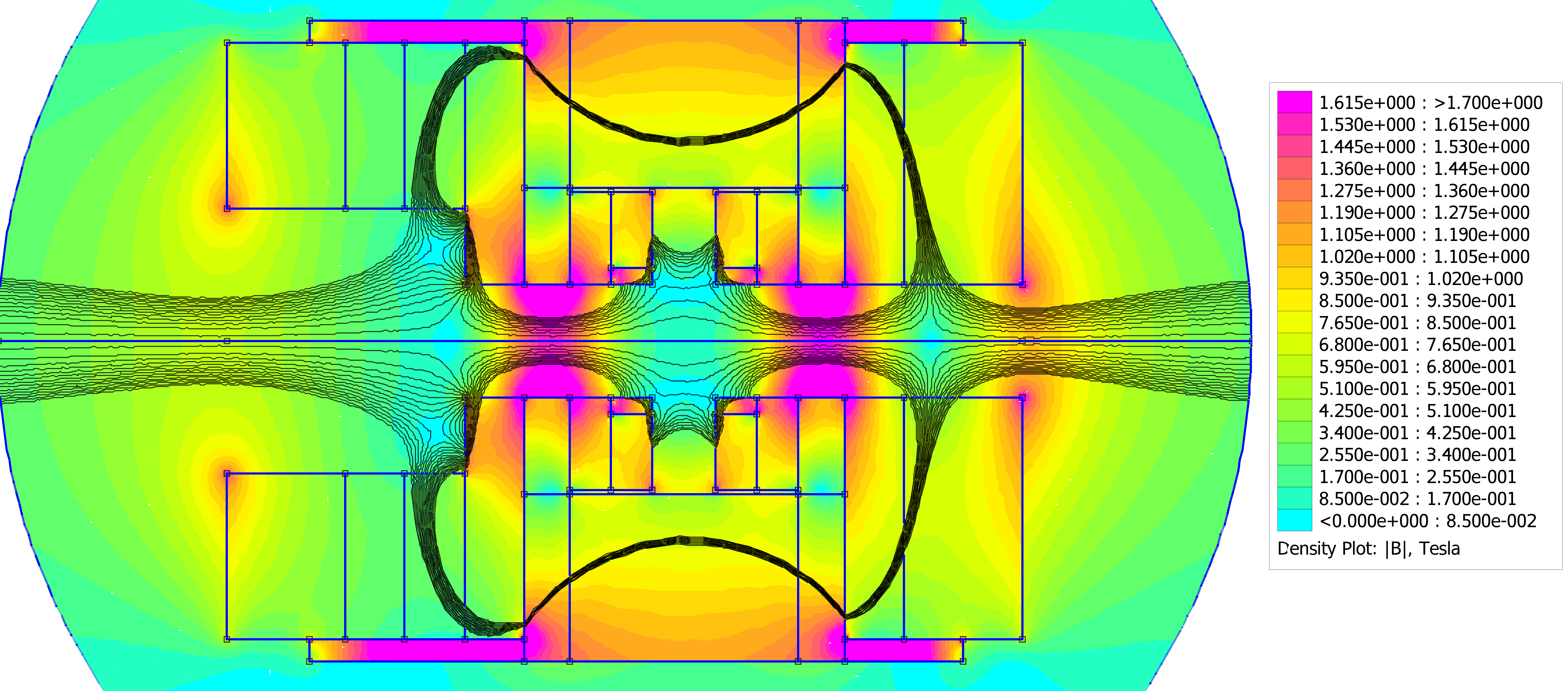}
\includegraphics[width=55mm]{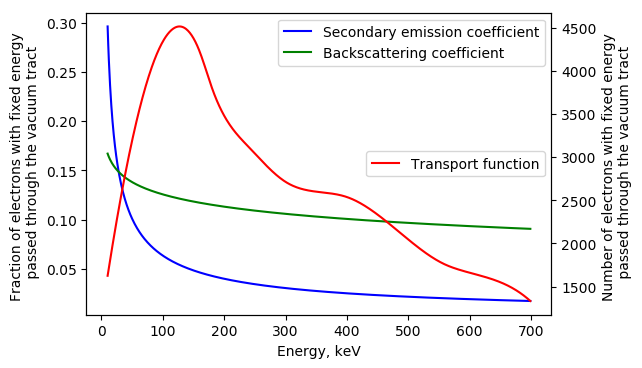}
\caption{\label{fig3} a) Magnetic field configuration of the GISMO facility; b) Factors that affected the electron signal.
	\vspace{0cm}}
\end{figure*}

Plasma confinement in GISMO is performed with the use of a simple mirror trap made of permanent magnets. The parameters of the trap are $ B_ {max} = 1.5 $ T, $B_ {min} = 0.25$ T, and the mirror ratio is, consequently, 6. The configuration of the magnets is such that there is a cusp, an area with $ B = 0 $, downstream the mirror trap (Fig.\ref{fig3}a). 

The ECR discharge is sustained by the gyrotron radiation with the frequency of $28$ GHz and power of up to $10$ kW. Such unique features, along with the small volume of the plasma chamber (approx. $50 cm^3$), make it possible to achieve record-breaking values of the specific energy input into the plasma of a continuous ECR discharge at a level of up to $ 50-100$  W/cm$^ 3 $, whereas for conventional ion sources it does not exceed $ 1-5 $ W/cm$^ 3 $.

The design of GISMO allows to to operate in a wide range of gas pressures in the discharge from $ 10 ^ {- 2}$ Torr  to $ 10 ^ {- 6}$ Torr, which makes it possible to conduct research both in classical (collisionless) and quasi-gasdynamic (collisional) confinement modes. During the experiments described below, the facility was operated in the range of pressures $0.1-3$ mTorr, where we got the most interesting results for our further research.

The energy distribution of electrons escaping the magnetic trap was obtained as follows. The method is similar to ion mass spectrometry with an inverted polarity of an electromagnet that deflects charged particles, as described in \cite{lit2}. Electrons that leave the plasma pass along the magnetic lines of the system, then consequently enter the diagnostic chamber, the magnetic analyzer and the secondary multiplier. The presence of a cusp in the system allows to observe only electrons that originated close to the symmetry axis of the system: thus, a small parallel electron beam was detected by the preamplifer. A significant part of the electrons is lost on the chamber walls, thus generating bremsstrahlung. A secondary electron multiplier and a Stanford SR570 current preamplifier were used to register electron currents as small as 1 nA. A voltage of -3.5 kV was applied to the cathode of the electron multiplier with respect to the chamber potential, which prevented the registration of electrons with energies below 3.5 keV. The design of the system made it possible to allow scanning with the resolution of about 1 keV, which was checked by extracting the electron beam with applying different values of the negative voltage. 

During the data processing for reconstructing the lost electron energy distribution (LEED), the following corrections for the electron current were taken into account (Fig. \ref{fig3}b). Firstly, the coefficients of secondary emission (\cite{lit3}) and backscattering (\cite{lit4}) of the electron multiplier cathode to correctly estimate the number of electrons that reached the detection system. Secondly, the transport function that determines the fraction of electrons with a particular energy lost on the way from the plasma electrode to the detector due to the geometric peculiarities of the system. The transport function was simulated by means of electron tracking in the static field, initial particle position was defined at the mirror plug with K-V distribution. It is important to note that the simulation conducted to obtain the transport function has a high computational cost, thus more accurate calculations will be acquired for the later research to get a more precise one.

Simultaneously with the LEED measurements, bremsstrahlung spectra were obtained. Amptek's XR-100T-CdTe X-ray detector was installed in a way to observe the end of the plasma chamber, where the energetic particles precipitated, causing bremsstrahlung. Despite the fact that plasma is also a source of bremsstrahlung, its density is orders of magnitude lower than the density of the solid walls. Thus, plasma bremsstrahlung is negligible when compared to a thick target one.

\section{Experimental results}
The LEEDs were obtained in a wide range of gas pressures, from $0.1$ mTorr to $3$ mTorr, and CW radiation powers of the gyrotron that supports the ECR discharge, from $1.3$ kW to $4.6$ kW. The operating gas was hydrogen. Energy distributions and bremsstrahlung spectra showed unconventional behavior as the function of external parameters.
\begin{figure*}[t]
\centering
\includegraphics[width=60mm]{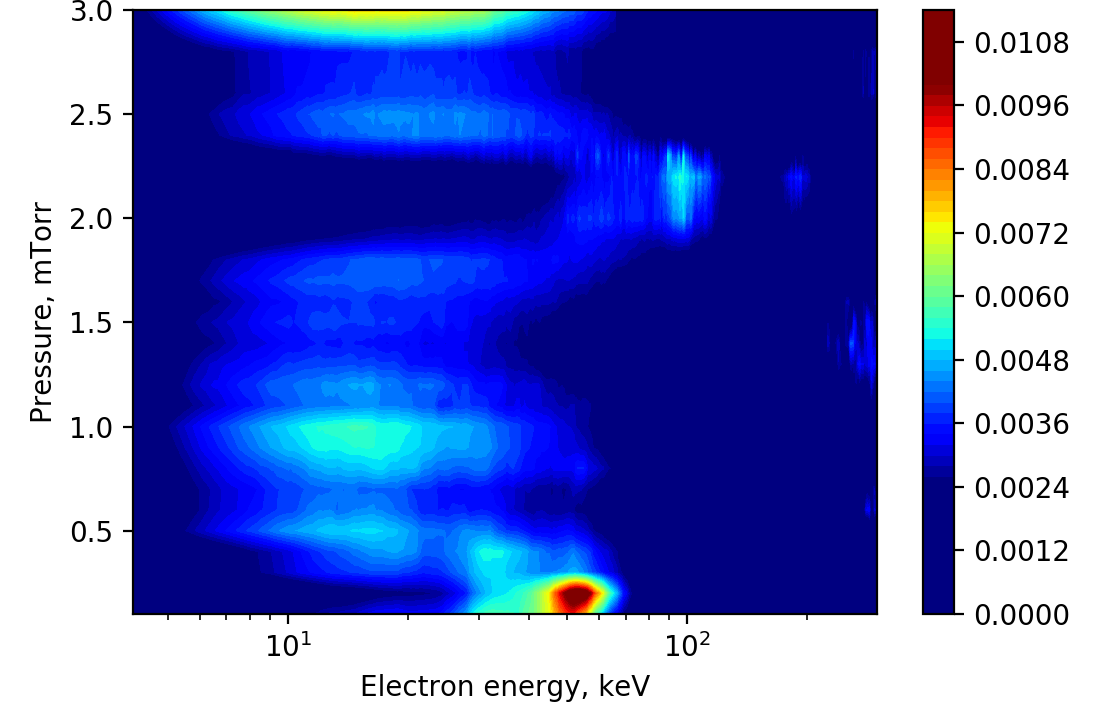}
\includegraphics[width=60mm]{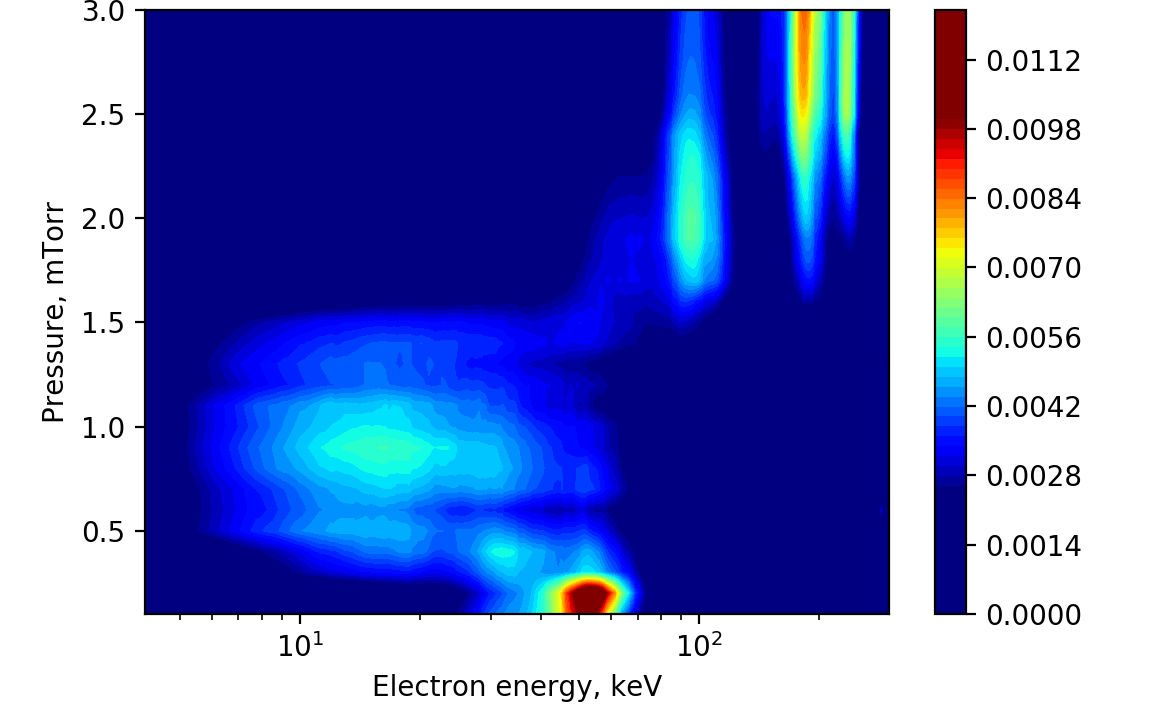}
\includegraphics[width=60mm]{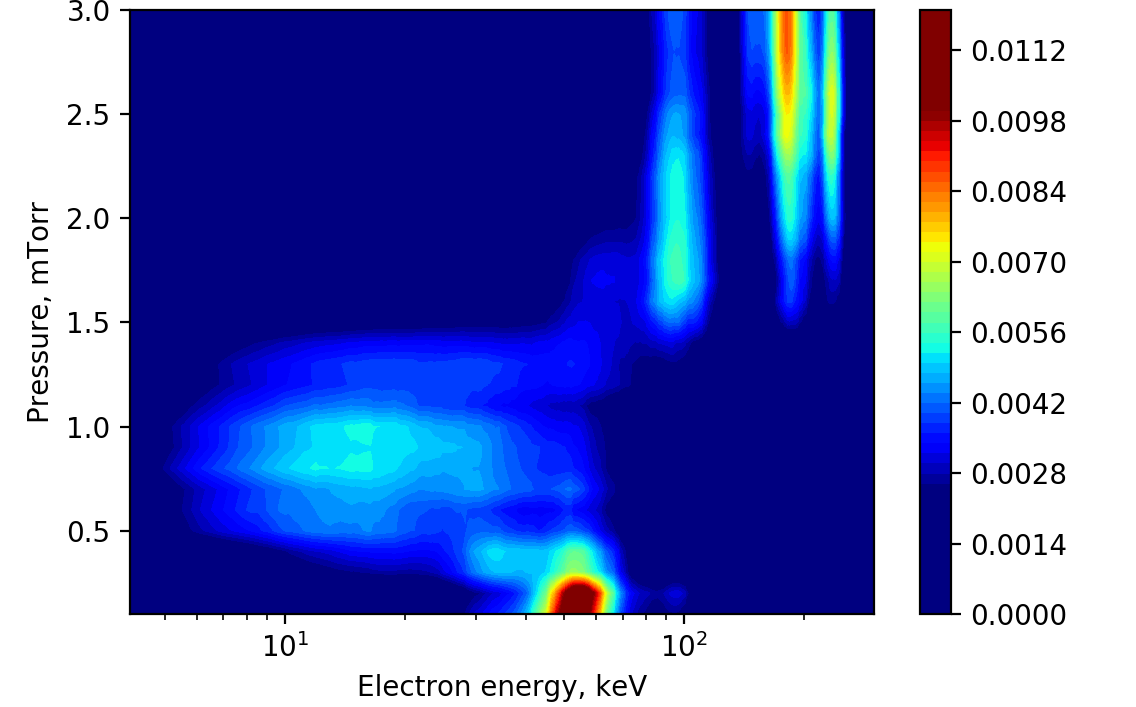}
\includegraphics[width=60mm]{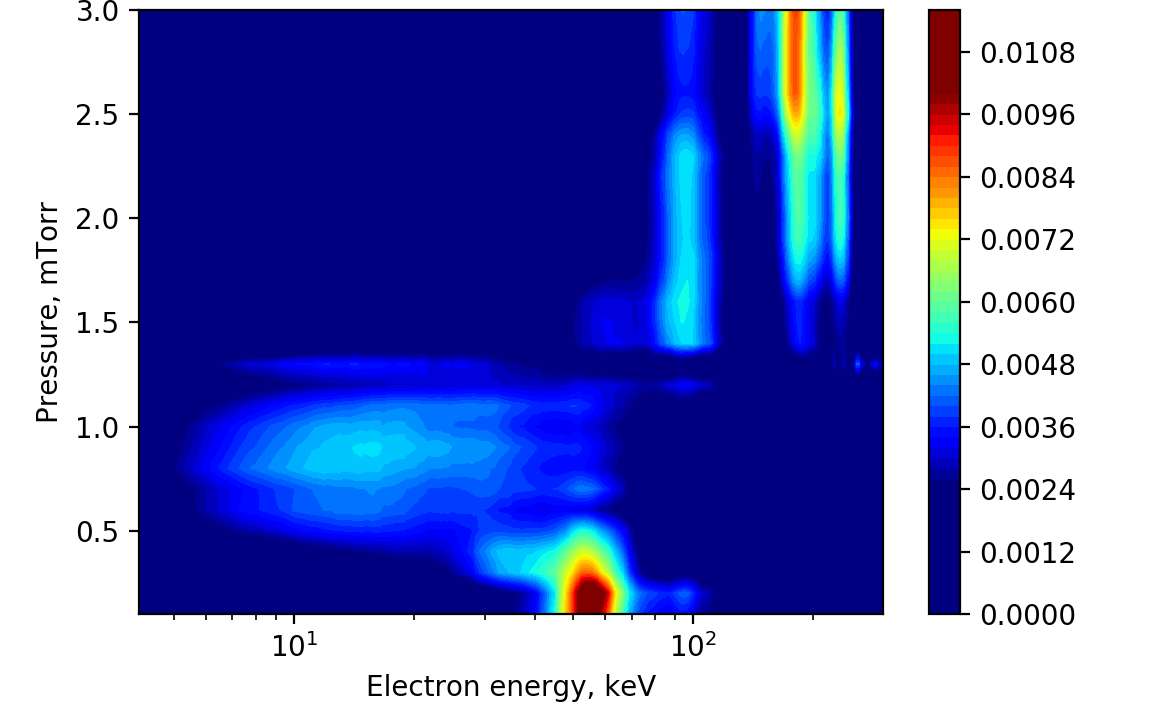}
\caption{\label{fig4} Lost electron energy distributions heatmaps measured for a) $1.7$; b) $2.3$; c) $3.3$; d)$4.6$ kW of gyrotron power
	\vspace{0cm}}
\end{figure*}

Figure \ref{fig4} shows the heatmaps of the LEED normalized per unit area for a large number of pressures at different powers. Here the x-axis is the electron energy, the y-axis is the pressure, and the color indicates the relative fraction of particles with a fixed energy at a fixed pressure. Such approach makes it possible to track the evolution of the shape of the electron distribution with the gas pressure.
It can be seen how strongly the distributions differ depending on the pressure of the neutral gas. At small ($0.1 - 0.2$ mTorr), the main fraction of particles has an energy of around $50-60$ keV and remains almost unchanged with an increase in the gyrotron power. At $0.3 - 1.7$ mTorr, the fraction of electrons is formed in a wide range of $5 - 30$ keV and weakly depends on the power. LEEDs at high pressures are characterized by a qualitative change in shape and the appearance of high-energy ($100 - 200$ keV) fractions of energetic electrons. It can be suggested that even relatively small changes in gas pressures can noticeably affect confinement conditions. Small values (< $0.5$ mTorr) correspond to nearly collisionless mode, which allows the development of the unstable hot electron fraction by interacting with microwave radiation not depending of the gyrotron power. However, it is clearly seen that when the power exceeds $1.7$ kW, the shape of the distribution abruptly changes for a big range of pressures (>$1.5 mTorr$).

\begin{figure*}[t]
\centering
\includegraphics[width=60mm]{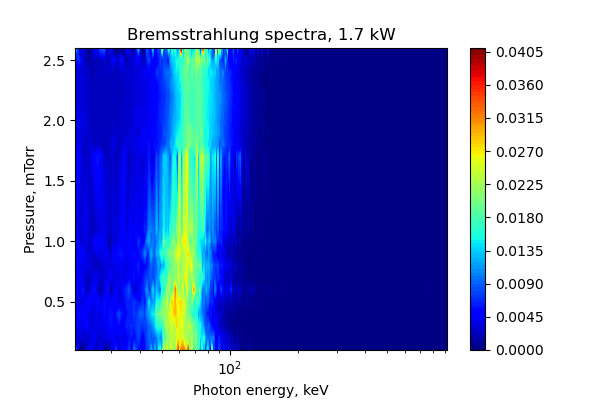}
\includegraphics[width=60mm]{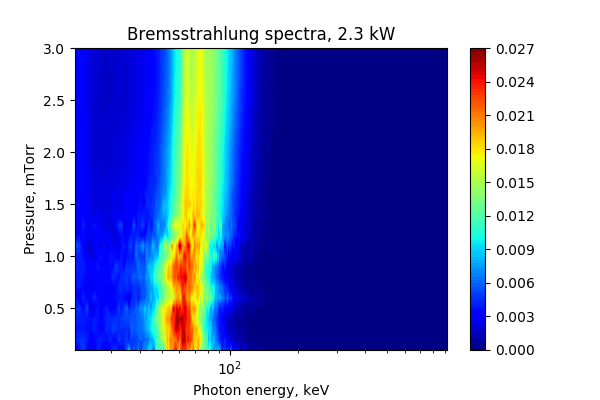}
\includegraphics[width=60mm]{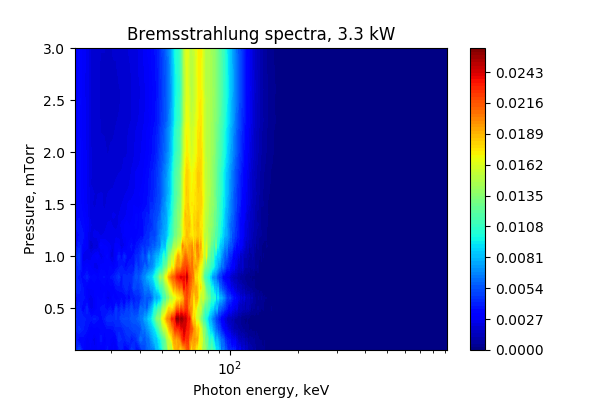}
\includegraphics[width=60mm]{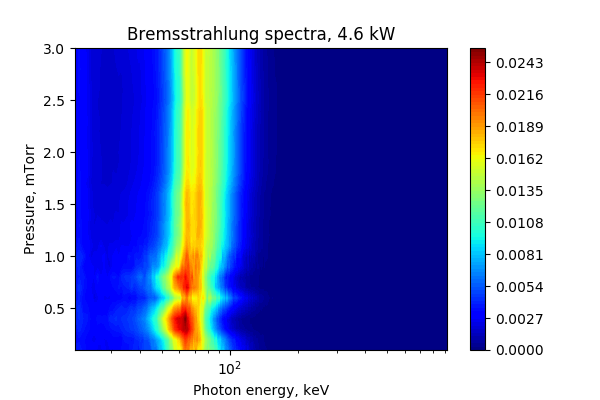}
\caption{\label{fig5} Bremsstrahlung spectra obtained for a) $1.7$; b) $2.3$; c) $3.3$; d)$4.6$ kW of gyrotron power
	\vspace{0cm}}
\end{figure*}

Corresponding bremsstrahlung spectra are shown in Fig. \ref{fig5} and barely differ with pressure. Some alterations are the result of a small number of quants received by the detector.

\begin{figure*}[t]
\centering
\includegraphics[width=70mm]{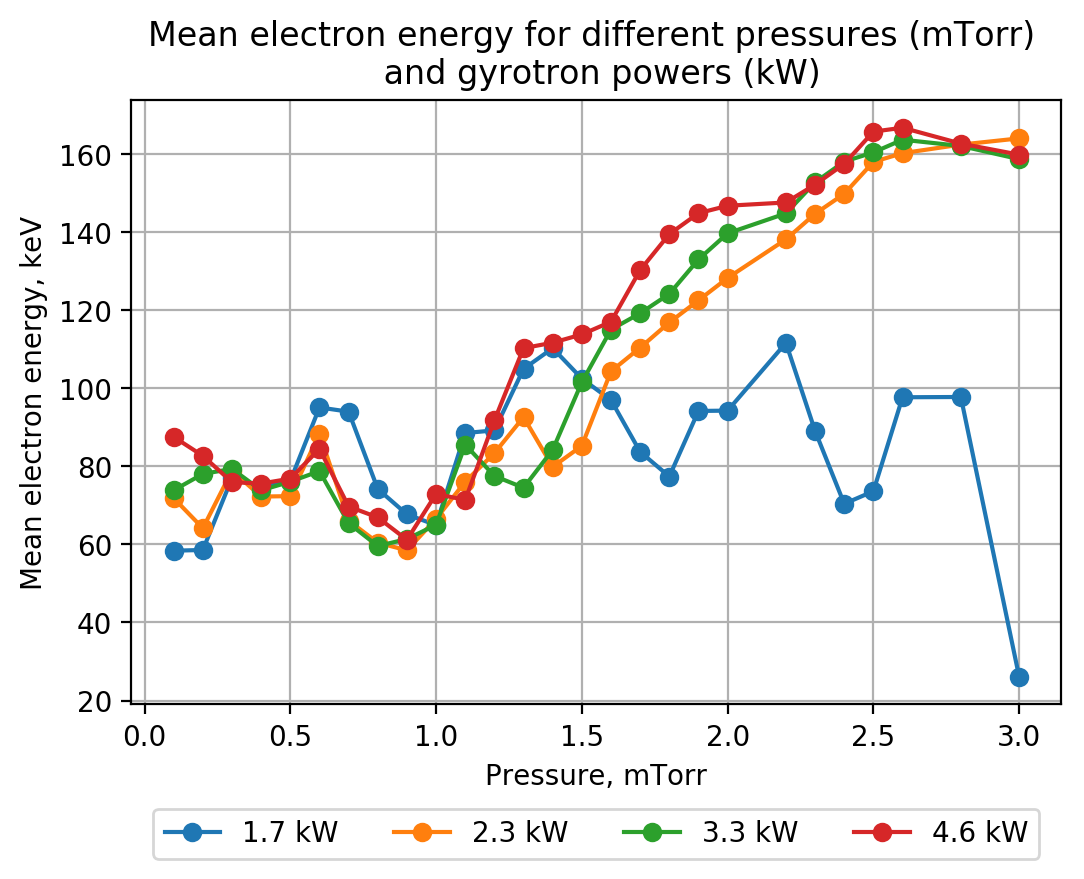}
\includegraphics[width=80mm]{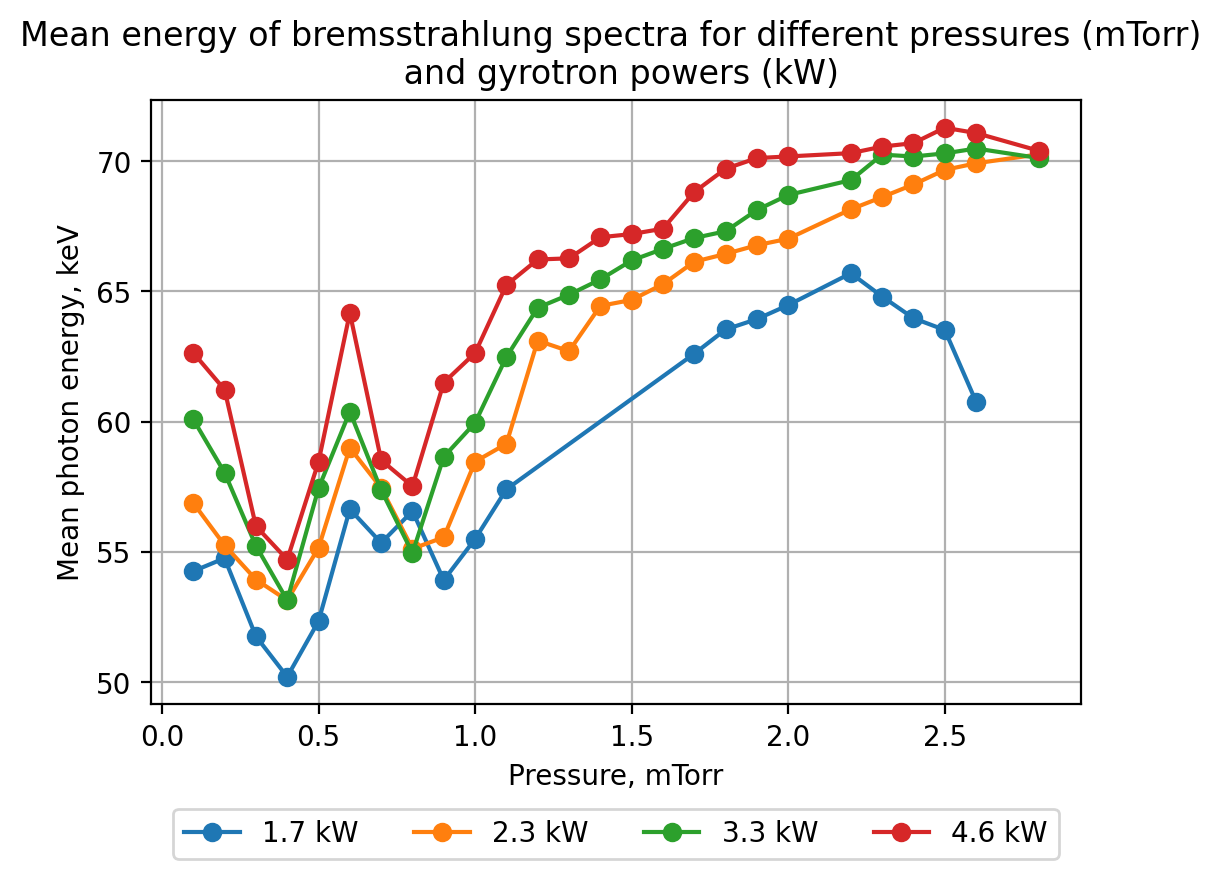}
\includegraphics[width=70mm]{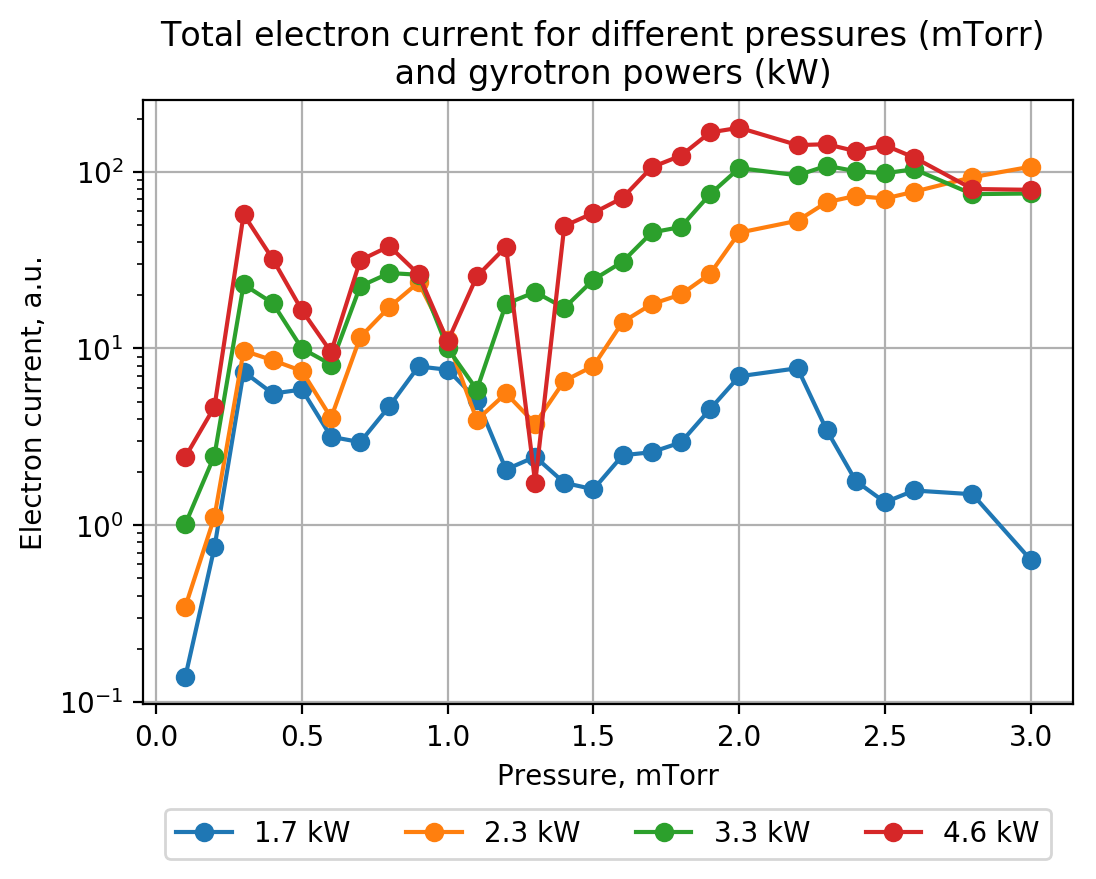}
\includegraphics[width=80mm]{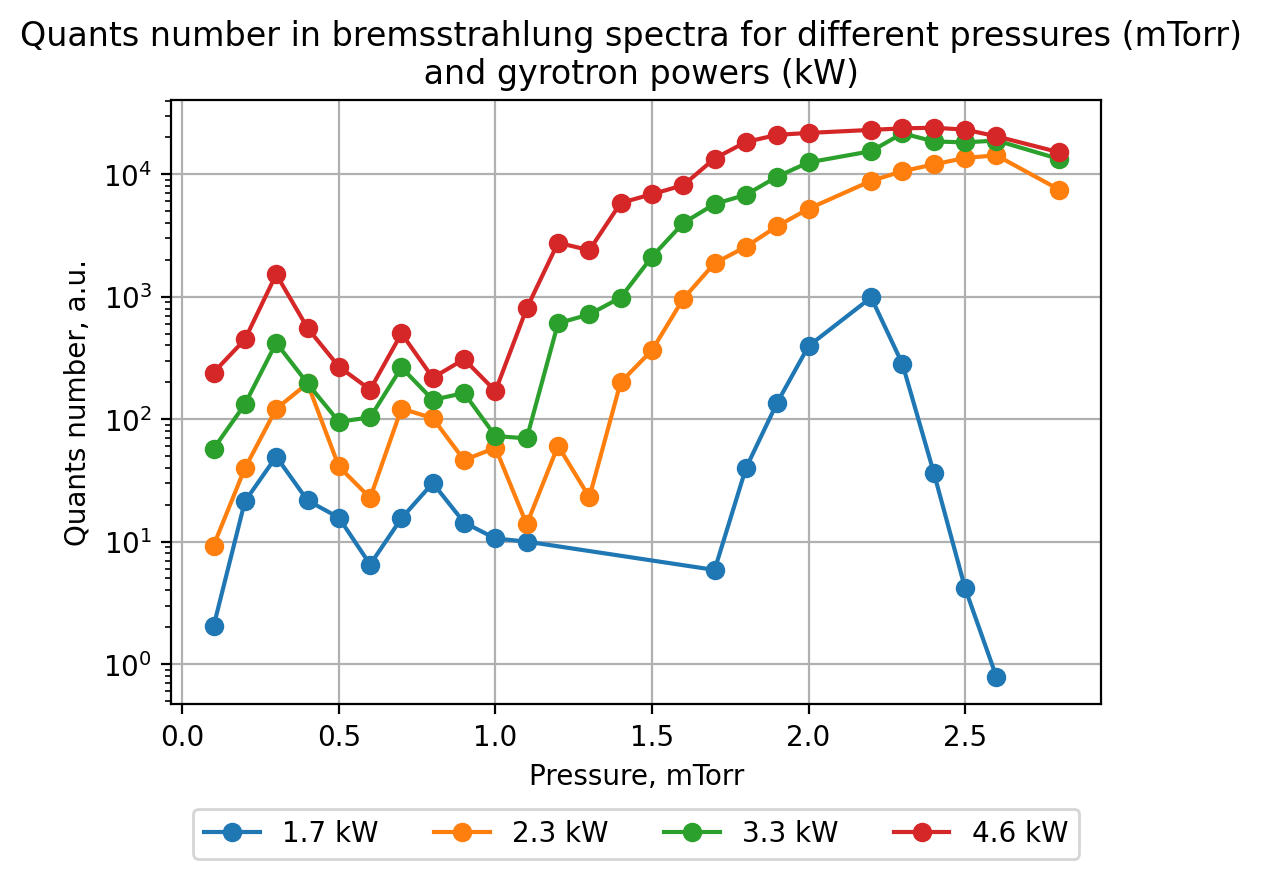}
\caption{\label{fig6} Integral LEED and bremsstrahlung characteristics depending on neutral gas pressure and gyrotron power, correspondingly: a), b) Mean energy; c) Total number of electrons; d) Input rate
	\vspace{0cm}}
\end{figure*}

However, heatmaps do not provide complete information for analysis, although they allow good visualisation of some effects. Firstly, it is difficult to understand the LEED shape out of them. Secondly, some changes in shape that are small in magnitude, but important in meaning can be unnoticed due to the peculiarities of color perception. At the same time, demonstration of all distributions in one two-dimensional plane also does not help in the interpretation of the data obtained and does not increase their information content. Moreover, it's important to add that Fig. \ref{fig4} shows the normalized LEED. Unnormalized distributions were even more difficult to analyze. One of the simplest solutions to this problem is the calculation of some integral characteristics of the distributions, namely, the total number of particles in the unnormalized distribution and the average energy. These data for LEED and bremsstrahlung are shown in Fig. \ref{fig6}. The average energy is taken as the mathematical average ($<\varepsilon>= \int \varepsilon f(\varepsilon)d\varepsilon$) over the distribution plotted in the given energy range, therefore this value is relative rather than absolute. As it was shown earlier, bremsstrahlung heatmaps do not provide any relevant information, but its integral characteristics seem to contain more sense for interpretation. 

Fig. \ref{fig6}a shows how the average energy of the distribution grows with increasing pressure for all power values, except for $1.7$ kW. This may be due to the fact that the probability of collisions and scattering into the loss cone for energetic electrons increases with pressure. In all distributions for $1.7$ kW, we detected fewer electrons (as seen in Fig. \ref{fig6}b) and the signal received on the oscilloscope could be influenced by the detector noise, which led to such oscillations in the values.

\begin{figure*}[t]
\centering
\includegraphics[width=60mm]{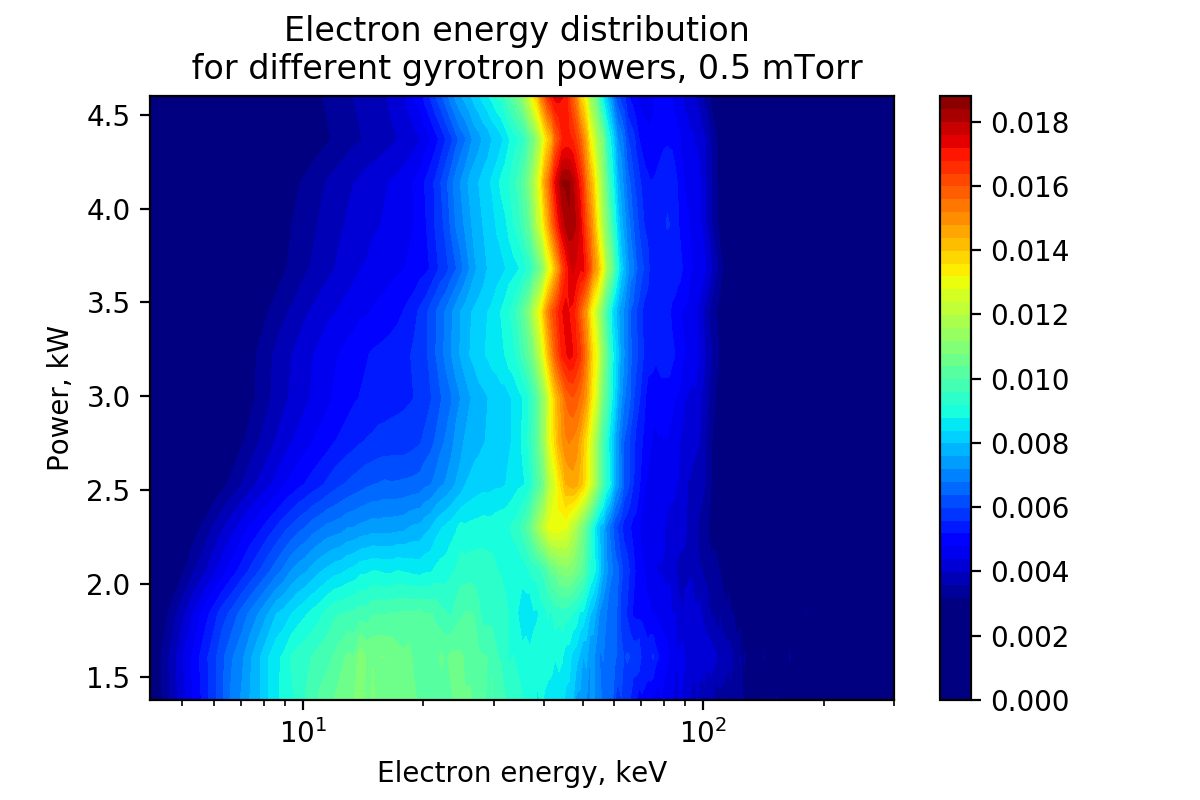}
\includegraphics[width=60mm]{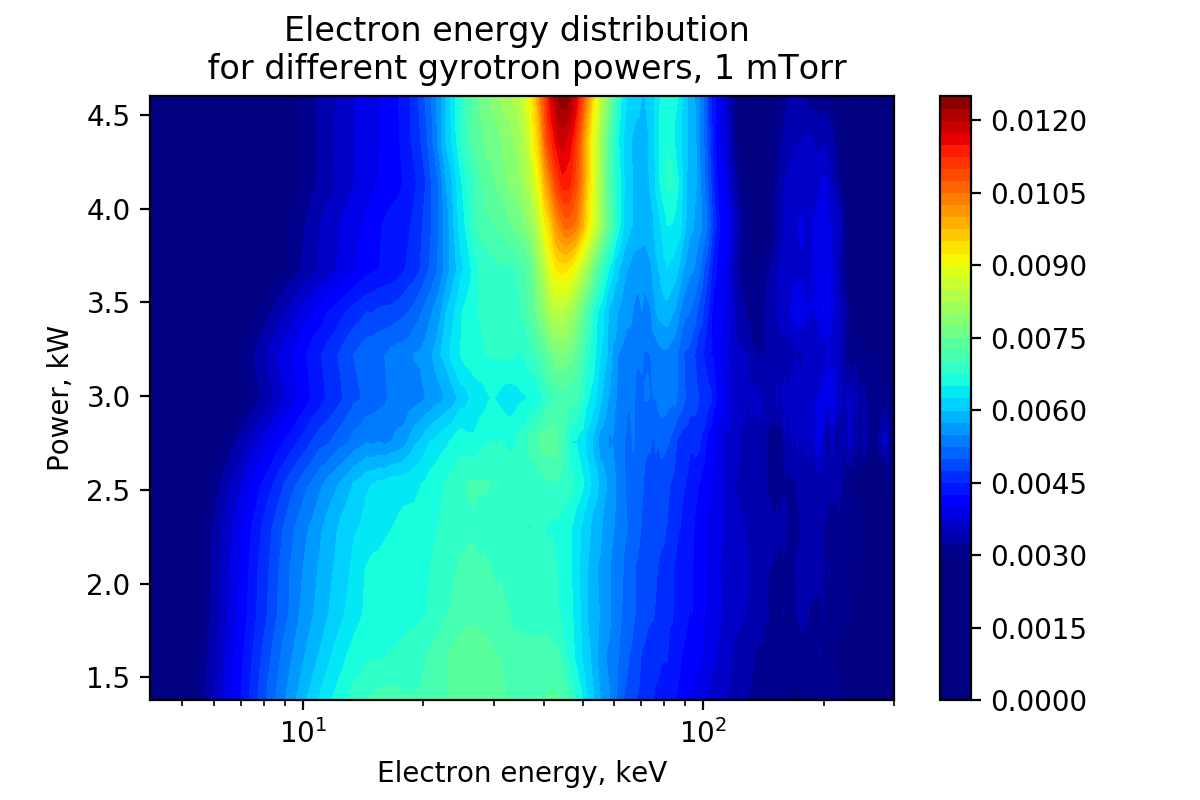}
\includegraphics[width=60mm]{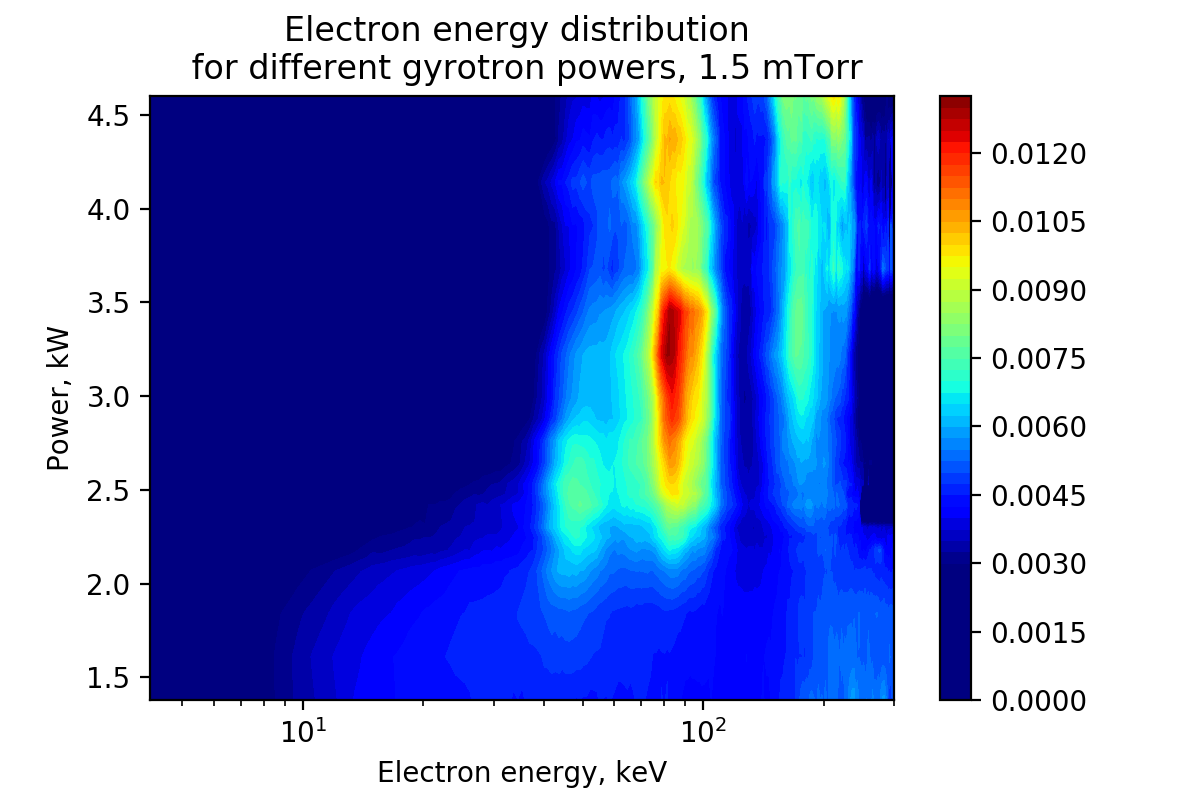}
\includegraphics[width=60mm]{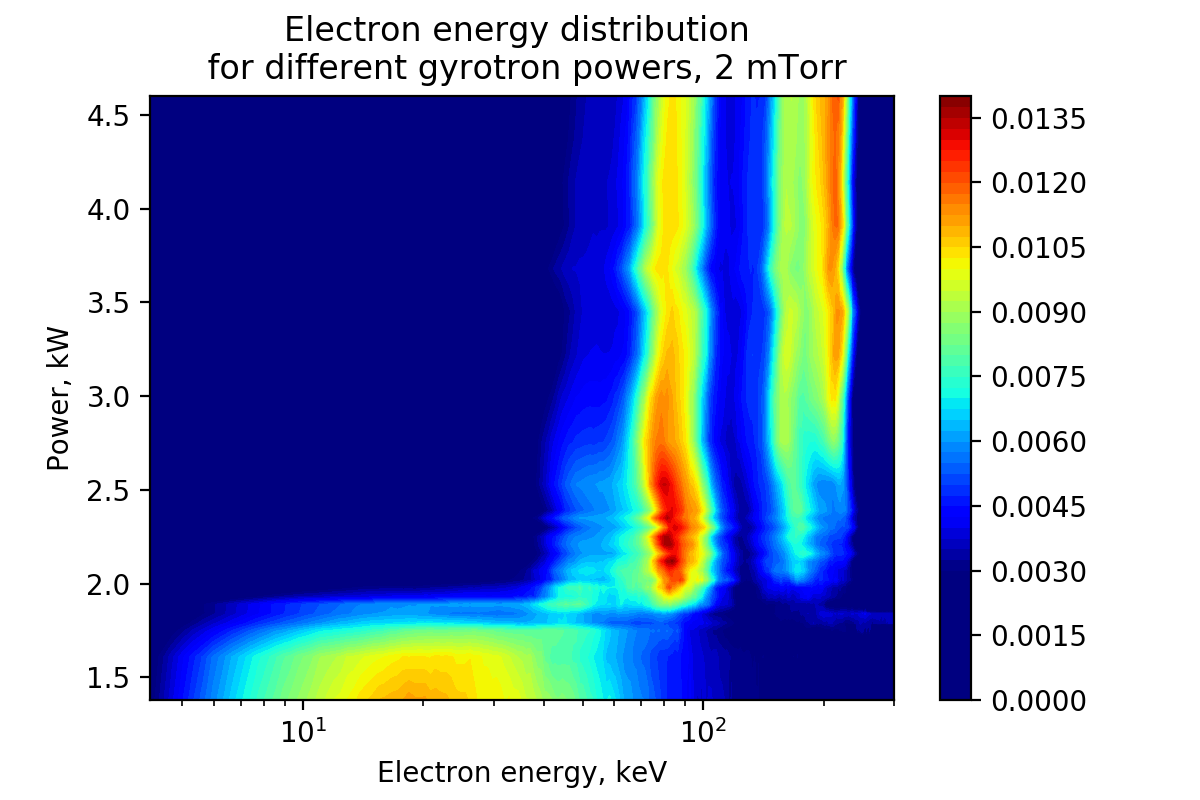}
\includegraphics[width=60mm]{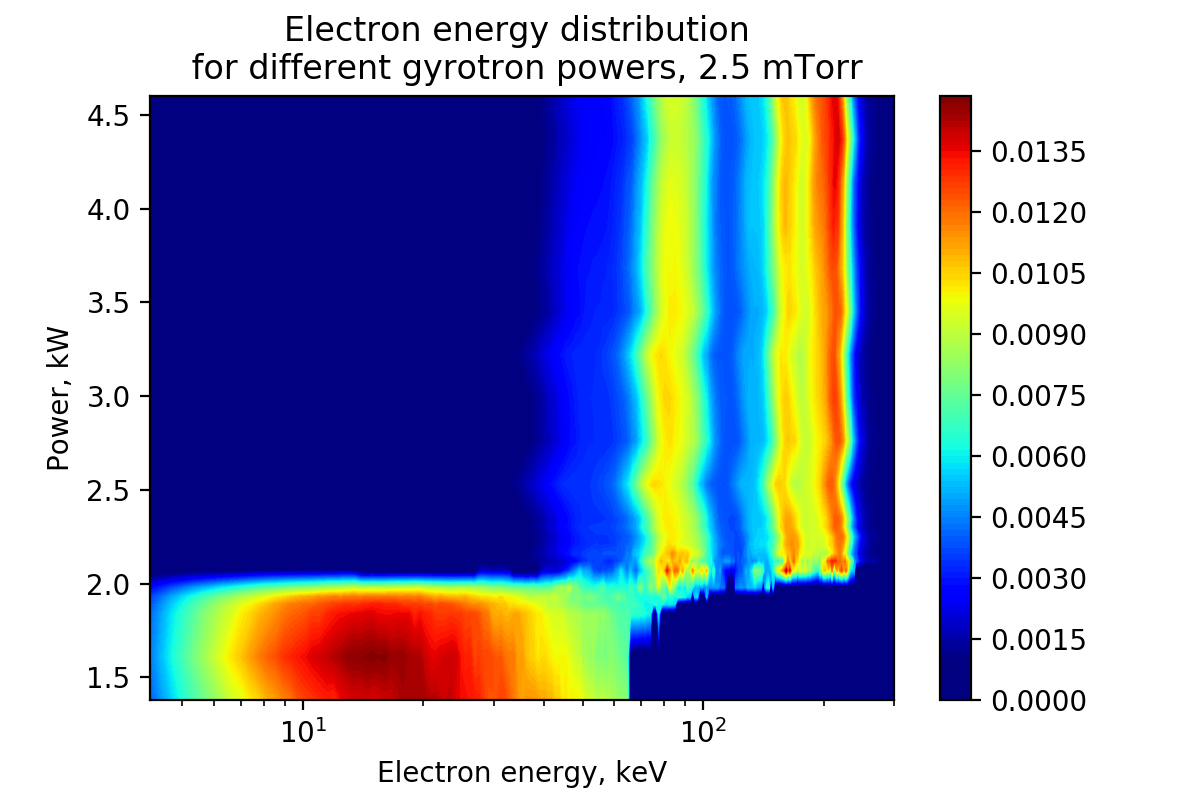}
\caption{\label{fig7} Lost electron energy distributions heatmaps measured for a) $0.5$; b) $1$ ; c) $1.5$; d)$2$; e) $2.5$ mTorr of neutral gas pressure
	\vspace{0cm}}
\end{figure*}

It is worth noting that at high powers the average energy is almost the same. This fact leads to the idea that inside the plasma, when a certain threshold power is exceeded, some processes take place that quantitatively and qualitatively change the energy distribution of particles; with a further increase in the power, the abrupt change is replaced by saturation: mean energy stays nearly the same, while total amount of electrons grows with a significantly lower rate. The evolution of the LEED with power with a small step for different pressures is considered in Fig. \ref{fig7}.

\begin{figure*}[t]
\centering
\includegraphics[width=70mm]{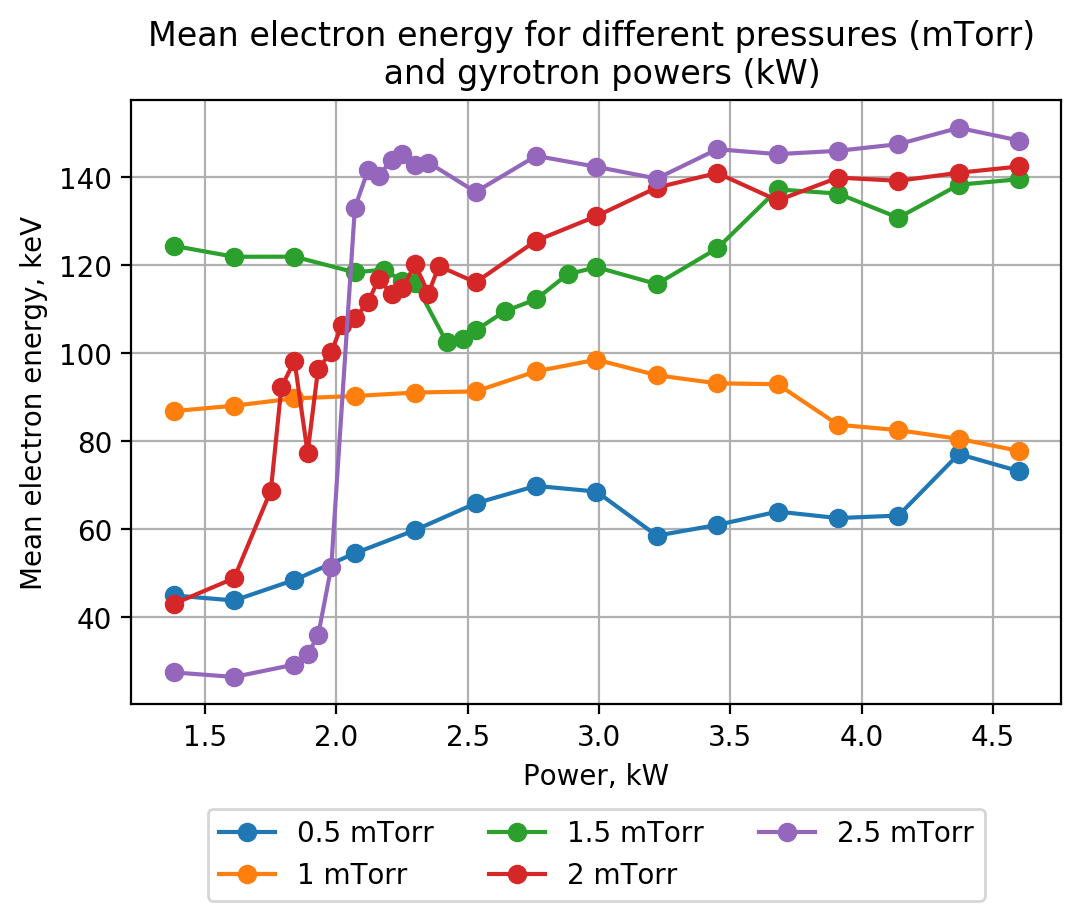}
\includegraphics[width=80mm]{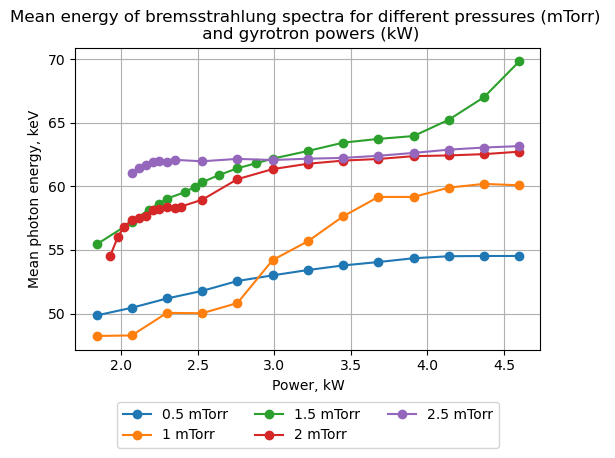}
\includegraphics[width=70mm]{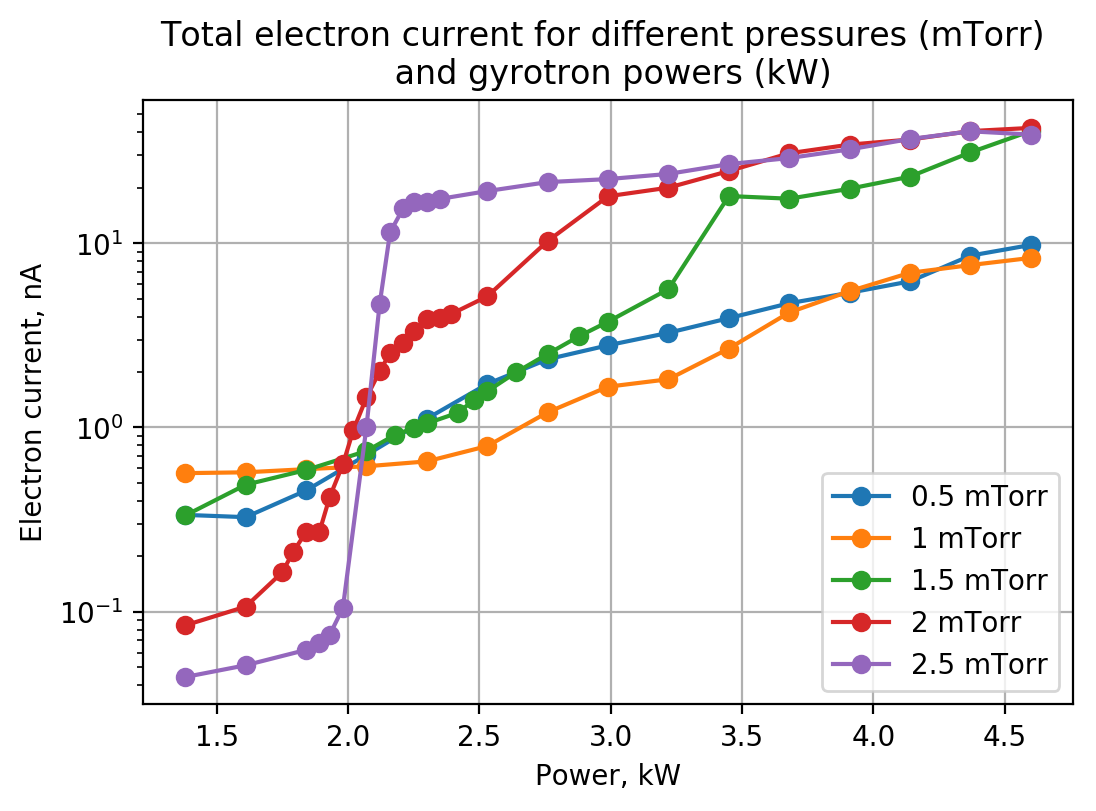}
\includegraphics[width=80mm]{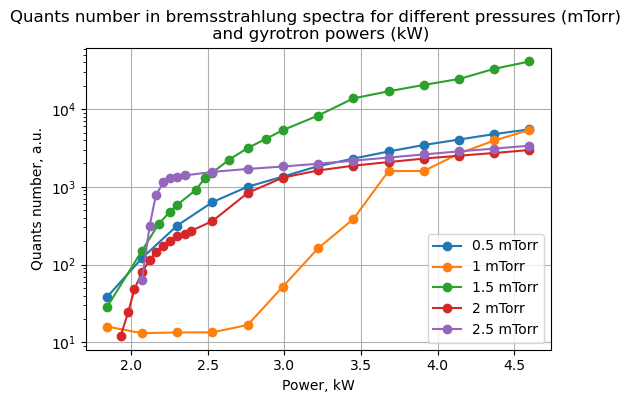}
\caption{\label{fig8} Integral LEED and bremsstrahlung characteristics depending on gyrotron power and neutral gas pressure, correspondingly: a), b) Mean energy; c) Total number of electrons; d) Input rate
	\vspace{0cm}}
\end{figure*}

A threshold-like effect is seen at the power of about $2$ kW, where the LEED shape dramatically changes along with the number of electrons reaching the detector and the intensity of bremsstrahlung (Fig. \ref{fig8}). A fraction of ‘hotter’ electrons is presumably formed from the instabilities. The latter increase of the gyrotron power barely changes the shape of the distributions, while electron current and bremsstrahlung count rate nearly reach saturation. However, the evolution of the LEED shape came along with bursts of electrons that formed the energetic peaks on the distribution. Such bursts can be associated with kinetic instabilities occurring in the ECR plasma when the gyrotron power exceeds some threshold value ($\propto 1.7$ kW).

The bremsstrahlung that is seen by detector through steel walls of the vacuum chamber ($>10-20$ keV) appears with the development of the unstable hot fraction of electrons. The spectrum shape does not change with the increase of the gyrotron power (Fig. \ref{fig7}), but the growth of the count rate is strongly correlated with the number of electrons escaping from the plasma. This fact leads to the conclusion that the main source of bremsstrahlung in the gasdynamic ECR ion sources might be hot electrons leaving the plasma as a result of the development of kinetic instabilities and hitting the walls of the chamber.

\section{Conclusion}
In this work, the energy distributions of the electrons lost from gyrotron-heated ECR plasma in an ion source with specific energy input of up to $100 \; W/cm^{3}$ were obtained for the first time. LEEDs were measured in a wide range of external parameters, such as gas pressure and gyrotron power. Obtained distributions, as it was anticipated, turned to be nontrivial functions that differ significantly from Maxwellian ones. The spectra of electron bremsstrahlung in the X-ray range and their dependence on the above parameters were also obtained. It was shown that the number of bremsstrahlung quanta and electrons noticeably increases in a certain pressure range. More importantly, the observations revealed that the shape of the LEED dramatically changes when the power exceeds certain threshold, presumably as a result of the development of the kinetic instabilities. This process quantitatively and qualitatively changes the characteristics of the plasma, at which high-energy peaks on the LEED, a large amount of bremsstrahlung and emitted electrons appear abruptly. With the latter increase in power, however, the average energies remain nearly constant. This gives the potential of the bremsstrahlung suppression by the EEDF modification and, consequently, optimization of the ion source performance.

After all, there still remains a question of the origin of the threshold. It can be suspected that some weak EM waves (e.g. cavity modes or plasma radiation) participate in the interaction processes as the additional 'heating'. Then, according to the theory described in the introduction, we saw the consequent processes of the inception of the unstable electron fraction and its suppression by the waves occurred simultaneously with it.

Additionally, there is another detail that is catching attention. In one of the works \cite{lit2a}, it was shown that the high-energy peak in the distribution function of emitted electrons is associated with the emission of one of the cavity modes excited by an unstable fraction of resonant electrons, which confirms the conclusions from the theory described in \cite{control}. The simultaneous measurements of LEED and a plasma microwave emission is the topic of the future research.

\section{Acknowledgements: } This research was supported by the grant of Russian Science Foundation (project number 19-12-00377). 

\end{document}